\documentstyle[twoside,fleqn,espcrc2,psfig]{article}

\newcommand{\AmS}{{\protect\the\textfont2
  A\kern-.1667em\lower.5ex\hbox{M}\kern-.125emS}}

\title{Two-Dimensional Dynamical Triangulation 
       using the Grand-canonical Ensemble}

\author{S. Oda\address{Nara Women's University,
	               Nara, Nara 630, Japan}
       ,
        N. Tsuda\address{High Energy Accelerator Research Organization (KEK),
                         Tsukuba, Ibaraki 305, Japan}
	and
	T. Yukawa\address{Coordination Center for Research and Education,
                          The Graduate University for Advanced Studies,\\
                          Hayama-cho, Miura-gun, Kanagawa 240-01, Japan}
                          $^{\! , \,\,{\rm b}}$
} 
       
\begin{document}

\begin{abstract}
The string susceptibility exponents of dynamically triangulated 
two dimensional surfaces with sphere and torus topology 
were calculated using the grand-canonical Monte Carlo method.
We also simulated the model coupled to d-Ising spins (d=1,2,3,5).
\end{abstract}

\maketitle

\section{Introduction}

For 2-dimensional quantum gravity 
the partition function with a fixed topology is approximated by
\begin{equation}
Z(A) \approx A^{\gamma-3} e^{\lambda_c A}
\end{equation}
for large $A$ limit,
where $A$ is the 2-dimensional volume (area),
$\lambda_c$ is the cosmological constant,
and $\gamma$ is the string susceptibility exponent.
The exponent $\gamma$ is known to be 
\begin{equation}
\gamma = \frac{1-h}{12} 
             \left[ c-25 - \sqrt{(25-c)(1-c)} \right] + 2 \label{eq:gamma}
\end{equation}
by analytical theories \cite{Liouville,Matrix_model},
where $h$ is the number of handles of the 2-dimensional surface,
and $c$ is the central charge of matter fields coupled to the surface.
When the dynamical triangulation (DT) recipe \cite{DT} is 
adopted to generate random surfaces,
it becomes possible to simulate the discretized 
2-dimensional gravity numerically.

In order to measure the exponent $\gamma$ with any topology, 
we introduce the grand-canonical Monte Carlo method \cite{Grand_Canonical}. 
In this method the finite size effect \cite{Finite_size_effects}
enters through the total area 
and becomes controllable in reasonably large size simulations.
Other method such as the MINBU (minimum neck baby universe) 
algorithm \cite{MINBU_sphere} 
have been successful in the measurement of $\gamma$ of sphere topology
for $c<1$.
However, in the MINBU algorithm 
measuring $\gamma$ for the case of a higher genus
is more complicated than for a sphere,
because the minimum neck baby universe must not be topologically non-trivial
and the baby universe distribution describing 
two topologically different $\gamma$ 
must be separated independently \cite{MINBU_torus}.
Eq.(\ref{eq:gamma}) shows that $\gamma$ becomes complex for $1<c<25$.
However, in a numerical simulation by dynamical triangulation 
the partition function is kept real for any value of $c$.
Then we evaluate the exponent $\gamma$ for $c>1$
using the grand-canonical Monte Carlo method.
\section{Model and Method}
\subsection{Model}
We start from the discretized Euclidean model.
The discretized partition function with a fixed topology is given by
\begin{equation}
Z(\lambda) = \sum_{T,N_2} W(T,N_2) e^{-\lambda N_2} Z_{\sigma},
                                                   \label{eq:partition_dis}
\end{equation}
where
\begin{equation}
Z_{\sigma} = \exp \left\{ \beta \sum_{\mu=1}^d \sum_{<ij>} 
                         \sigma_i^{\mu} \sigma_j^{\mu}         \right\}. 
\end{equation}
Here $W(T,N_2)$ is the symmetry factor, 
$T$ refers to a specific triangulation, 
$N_i$ is the total number of $i$-simplices,
$\lambda$ is a parameter,
and $\beta$ is a coupling strength of Ising spins.
We adopt the model excluded ``tadpole'' and ``self-energy'' diagrams.
There are $d$ kinds of Ising spins $\sigma^{\mu}$ on a vertex,
which carry $c=\frac{d}{2}$ central charge in total.

\subsection{Measurement of $\gamma$}
For the large-$N_2$ limit the partition function is written as
\begin{equation}
Z(\lambda) \approx \sum_{N_2} N_2^{\gamma -3}e^{-(\lambda - \lambda_c) N_2}.
\end{equation}
Then, the distribution $P(N_2)$ is given by 
$P(N_2) \propto N_2^{\gamma -3}e^{-(\lambda-\lambda_c)N_2}$.
From
\begin{equation}
\frac{\ln P(N_2)}{\ln N_2} = (\gamma -3) 
                       - (\lambda-\lambda_c)\frac{N_2}{\ln N_2} + const,
\end{equation}
if we tune $\lambda$ close to $\lambda_c$ within the precision
$ | \lambda - \lambda_c | \ll \frac{\ln N_2}{N_2} $ ,
we can evaluate the exponent $\gamma$ by plotting $\ln P(N_2)$
as a function of $\ln N_2$ within the region 
$N_2^{min} \leq N_2 \leq N_2^{max}$.

\subsection{Grand-canonical simulation}
We explain the method used to generate DT surfaces in the case of 
pure gravity ($c=0$).
Suppose that we wish to increase the number of real vertices 
from $N_0$ to $N_0+1$; to do this
we choose two dual links belonging to the same dual loop $L_i$, and
insert a ``propagator'' dual link joining the two dual links,
thus creating an additional dual loop $L_{N_0 +1}$.
For the inverse move, $N_0 \rightarrow N_0 -1$, we choose a dual link at random,
and remove it, 
resulting in two dual loops $L_i$ separated by the dual link merging into one.
The numbers of possible increase and decrease moves
from configuration $A$ are
\begin{equation}
n_A^{inc} = \sum_{i=1}^{N_0} \frac{q_i(q_i-1)}{2} , \;
n_A^{dec} = \sum_{i=1}^{N_0} \frac{q_i}{2} ,
\end{equation}
where $\{ q_i \}$ is the set of coordination numbers of a configuration $A$.

In these moves the ergodic property is automatically satisfied.
However, we must treat the detailed valance carefully \cite{Detail_bal}, 
since the total number of possible moves depends on 
$N_0$ and  $\{ q_i \}$.
The detailed valance equation, that the move should satisfy,
between configurations $A$ and $B$, is given by
\begin{equation}
\frac{W_A}{n_A} P_{A \rightarrow B} =
\frac{W_B}{n_B} P_{B \rightarrow A},              \label{eq:detail_bal}
\end{equation}
where $P_{A \rightarrow B}$ is the transition probability from configurations
$A$ to $B$, $W_A$ is the Boltzmann factor determined 
from eq.(\ref{eq:partition_dis}), 
and $n_A$ is the total number of possible moves from configuration $A$.

For the case $c=\frac{d}{2}$, we can easily extend the method by modifying
the total number of moves for including the freedom of spins as
\begin{equation}
n_A = \sum_{i=1}^{N_0}\left[ \frac{q_i(q_i-1)}{2} \cdot 4^d 
                           + \frac{q_i}{2} \cdot 2^d          \right].
\end{equation}
For the spin variables, Monte Carlo trials with Wolff's cluster algorithm 
\cite{Cluster}  are performed after some geometrical moves.

\section{Results and Discussions}

\begin{table*}[hbt]
\setlength{\tabcolsep}{1.3pc}
\newlength{\digitwidth} \settowidth{\digitwidth}{\rm 0}

\catcode`?=\active \def?{\kern\digitwidth}
\caption{String susceptibility exponent $\gamma$ of sphere and torus}
\label{Table:gamma}
\begin{tabular*}{\textwidth}{@{}lrrrrrrr@{}}
\hline
        & \multicolumn{3}{c}{Sphere ($h=0$)}
        & 
        & \multicolumn{3}{c}{Torus ($h=1$)} \\
\cline{2-4} \cline{6-8}
        & \multicolumn{1}{c}{$\gamma$} 
        & \multicolumn{1}{c}{$\lambda_c$} 
        & \multicolumn{1}{c}{$\beta_c$}
        &
        & \multicolumn{1}{c}{$\gamma$} 
        & \multicolumn{1}{c}{$\lambda_c$} 
        & \multicolumn{1}{c}{$\beta_c$}         \\
\hline
$c=0.0$  & $-0.517(27)$ & $1.12467$ & ------  & & $2.014(24)$ & $1.12469$ & -------- \\
$c=0.5$  & $-0.395(27)$ & $1.53716$ & $0.227$ & & $1.995(23)$ & $1.52890$ & $0.2163$ \\
$c=1.0$  & $-0.009(26)$ & $1.94848$ & $0.226$ & & $1.971(22)$ & $1.93335$ & $0.2163$ \\
$c=1.5$  & $ 0.112(27)$ & $2.35843$ & $0.225$ & & $1.980(23)$ & $2.33841$ & $0.2163$ \\
$c=2.5$  & $ 0.238(27)$ & $3.17513$ & $0.223$ & & $2.018(25)$ & $3.14950$ & $0.2163$ \\
\hline
\end{tabular*}
\end{table*}

We measured the string susceptibility exponent $\gamma$ 
of the 2-dimensional surface ($h=0,1$)
coupled to $d$-Ising spins ($d=0,1,2,3,5$)
using the grand-canonical Monte Carlo method.
When Ising spins are put on the surface,
we determine the critical temperature $\beta_c$ 
from the peak of magnetic susceptibility 
for each combination of $h$ and $c$.
Especially in the case of a torus ($h=1$) the experimental values $\beta_c$ 
were quite close to
the theoretical value ($\beta_c \approx 0.2163$).
One sweep is defined by $2000 \times 2^d$ trial moves 
and 5 cluster operations of Ising spin.

In order to obtain $P(N_2)$, we measure the number of triangles $N_2$ 
$10^5$ times every 5 sweeps within the region $1000 \leq N_2 \leq 3000$.
Our results are summarized in Table \ref{Table:gamma}.
The errors indicated in parentheses for the last two digits are only 
statistical errors estimated from the least squares fits.
The systematic errors are 
estimated to be around $0.03 \sim 0.02$.

\begin{figure}[htb]
\centerline{\psfig{file=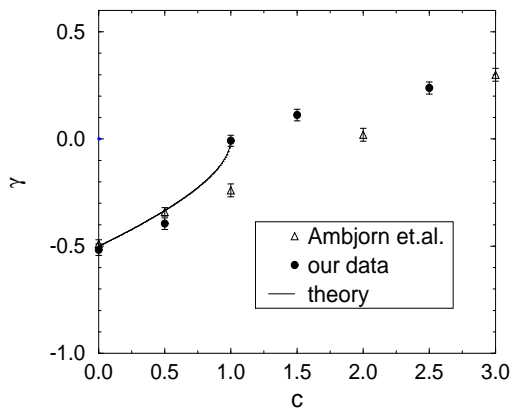,height=6cm,width=7cm}} 
\vspace{-1cm}
\caption{String susceptibility exponents $\gamma$'s of
a sphere with various central charge $c$'s.
Solid line represents eq.(2) with $h=0$.}
\label{fig:sphere}
\end{figure}

\begin{figure}[htb]
\centerline{\psfig{file=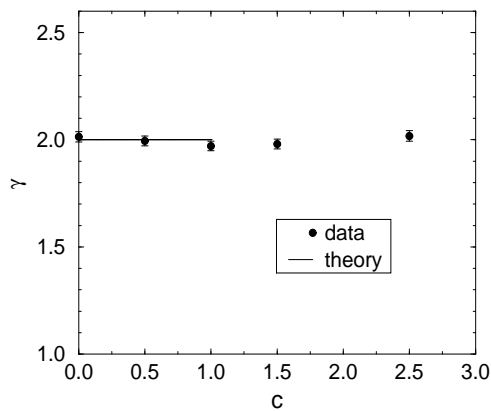,height=6cm,width=7cm}} 
\vspace{-1cm}
\caption{String susceptibility exponents $\gamma$'s of
a torus with various central charge $c$'s.
Solid line represents eq.(2) with $h=1$.}
\label{fig:torus}
\end{figure}

For the pure gravity ($c=0$), 
Table \ref{Table:gamma} shows that both the exponent $\gamma$ and
the cosmological constant $\lambda_c$ are good agreements 
to the theoretical values ($\gamma=-0.5$, 
$\lambda_c = \frac{1}{2} \ln \frac{256}{27} \approx 1.12467 $).
Furthermore, in Figs.\ref{fig:sphere},\ref{fig:torus}
the exponents $\gamma$'s are reproduced closely to the theoretical predictions
as long as $c \leq 1$.
This proves our expect that 
the grand-canonical method is almost free from the finite size effects.
Since we could check that our method is able to treat torus topology,
it is easy to expand to the case of a higher genus.

\begin{center}
{\large ---Acknowledgements---}
\end{center}
It is a pleasure to thank H.Kawai for many helpful discussions.
Two of the authors (N.T. and S.O.) are supported by Research Fellowships of the 
Japan Society for the Promotion of Science for Young Scientists.

\vspace{-1cm}
\vspace{-1cm}


\begin{thebibliography}{99}
\bibitem{Liouville} 
V. G. Knizhnik, A. M. Polyakov and A. B. Zamolodchikov, 
Mod. Phys. Lett. A3 (1988) 819; 
J. Distler and H. Kawai, 
Nucl. Phys. B321 (1989) 509;
F. David, Mod. Phys. Lett. A3 (1988) 1651.
\bibitem{Matrix_model}
E. Br\'{e}zin and V. Kazakov, 
Phys. Lett. B236 (1990) 144;
M. Douglas and S. Shenker, 
Nucl. Phys. B335 (1990) 635; 
D. J. Gross and A. A. Migdal, 
Phys. Rev. Lett. 64 (1990) 717.
\bibitem{DT}
V. A. Kazakov, I. K. Kostov and A. A. Migdal, Phys. Lett. B157 (1985) 295; 
J. Ambj\o rn, B. Durhuus and J. Fr$\ddot{o}$hlich, 
Nucl. Phys. B257 [FS14] (1985) 433;
F. David, 
Nucl. Phys. B257 [FS14] (1985) 543.
\bibitem{Grand_Canonical}
J. Jurkiewicz, A. Krzywicki and B. Petersson, Phys. Lett. B177 (1986) 89;
J. P. Kownacki and A. Krzywicki, Phys. Rev. D50 (1994) 5329.
\bibitem{Finite_size_effects}
N. D. Hari Dass, B. E. Hanlon and T. Yukawa, Phys. Lett. B368 (1996) 55.
\bibitem{MINBU_sphere}
S. Jain and S. D. Mathur, Phys. Lett. B286 (1992) 239.
\bibitem{MINBU_torus}
G. Thorleifsson and S. Catterall, Nucl. Phys. B461 (1996) 350.
\bibitem{Detail_bal}
N. Tsuda, A. Fujitsu and T. Yukawa,
Comp. Phys. Comm. 87 (1995) 372.
\bibitem{Cluster}
U. Wolff, Phys. Rev. Lett. 62 (1989) 361.
\end{thebibliography}
\end{document}